\documentstyle[aps,epsf,eqsecnum]{revtex}
\RequirePackage{bm}
\newcommand{\nnb}{\nonumber}
\newcommand{\V}{\nabla\!}
\newcommand{\ppp}{\partial}
\newcommand{\varphiMuNu}[1]{{\cal L}_{\mu\nu}\varphi^{#1}}
\renewcommand{\~}{\tilde}

\begin{document}
%======================================%
%<<<<<<<<<<<< TITLE PAGE >>>>>>>>>>>>>>%
%======================================%

%
% Osaka University Heading
%
\leftline{\epsfbox{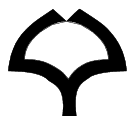}}
\vspace{-10.0mm}
\thispagestyle{empty}
{\baselineskip-4pt
 \font\yitp=cmmib10 scaled\magstep2
 \font\elevenmib=cmmib10 scaled\magstep1 \skewchar\elevenmib='177
 \leftline{\baselineskip20pt
           \hspace{12mm} % for revtex
           \vbox to0pt
             { {\yitp\hbox{Osaka \hspace{1.5mm} University} }
               {\large\sl\hbox{{Theoretical Astrophysics}} }
              \vss}
          }
}
%
% Preprint numbers
%
{\baselineskip0pt
 \rightline{\large\baselineskip14pt\rm\vbox
            to20pt{\hbox{OU-TAP-170}
                   \hbox{YITP-01-94}
              \vss}
           }
}
\vskip15mm
%
% Title and Author
%
\begin{center}
{\large\bf Quantum Radion on de Sitter branes}
\end{center}
\begin{center}
{\large
   Uchida Gen$^{1,2}$ and  Misao Sasaki$^{1}$\\
\bigskip
\small{\sl{  $^{1}$ Department of Earth and Space Science, Graduate School 
of Science,\\
          Osaka University, Toyonaka 560-0043, Japan\\
              \vskip1.5mm    and\\  \vskip1.5mm
     $^{2}$ Yukawa Institute for Theoretical Physics, Kyoto University,
      Kyoto 606-8502, Japan
           }
       }
}
\end{center}
%======================================%
%<<<<<<<<<<<<< ABSTRACT >>>>>>>>>>>>>>>%
%======================================%
\begin{abstract}
The quantum fluctuation of the relative location of two
$(n-1)$-dimensional de~Sitter branes (i.e., of $n$ spacetime dimensions)
embedded in the $(n+1)$-dimensional anti-de Sitter bulk, which we shall
call the quantum radion, is investigated at the linear perturbation level.
The quantization of the radion is done by deriving the
effective action of the radion. Assuming the positive tension brane
is our universe, the effect of the quantum radion is
evaluated by using the effective Einstein equations on the brane
in which the radion contributes to the effective energy momentum
tensor at the linear order of the radion amplitude.
Specifically, the rms effective energy density arising from
the quantum radion is compared with the background
energy density. It is found out that this ratio remains small
for reasonable values of the parameters of the model
even without introducing a stabilizing mechanism for radion,
although the radion itself has a negative mass squared and is unstable.
The reason behind this phenomenon is also discussed.
\end{abstract}


%======================================%
%<<<<<<<<<<<<<   MAIN   >>>>>>>>>>>>>>>%
%======================================%

\section{Introduction}

Based on the idea of a brane-world suggested from string
theory\cite{Antoniadis},
Randall and Sundrum proposed an interesting scenario that we may live
on either of two boundary 3-branes with positive and negative tensions
in the 5-dimensional anti-de Sitter space (AdS) \cite{RS1,RS2}.
One of the attractive features of the Randall-Sundrum (RS) scenario
is that the gravity on the brane is confined within a short distance
from the brane even for an infinitely large extra-dimension\cite{RS2,GT}.
This applies to the positive tension brane
and it is because the AdS bulk on both sides of the positive tension
brane shrinks exponentially as one goes away from the brane.

Since the RS scenario gives an exciting, new picture of our universe,
it is clearly important to study various aspects of this scenario
to test it or to give constraints on its parameters.
One good example is an analysis done by Garriga and Tanaka\cite{GS},
in which they have shown that the radion in the original RS scenario
acts as a Brans-Dicke scalar on the branes at the
linear perturbation order and the effective gravity is that of a
Brans-Dicke theory with positive and negative
Brans-Dicke parameters on the positive and negative tension
branes, respectively, where the values of the Brans-Dicke parameters
are determined by the distance between the two branes.
In a previous paper, we have shown that essentially the same
situation arises in the case of two de Sitter (dS) branes
embedded in the AdS bulk. However, we have also shown that the radion
effectively has a negative mass squared with its absolute value
proportional to the curvature of the dS brane, hence is unstable
if it can fluctuate by itself without the matter
energy momentum tensor.

The phenomenon we shall study in this paper is the quantum
fluctuation of this mode, called the quantum radion.
To make clear what we mean by the quantum radion, let us
describe the mode in more detail.
Our brane universe may be displaced from the 0th order trajectory of a
homogeneous and isotropic brane. By appropriately fixing the coordinate
gauge, the displacement perpendicular to the brane can be described by
a scalar function on the brane.
Assuming there are two branes that are
fixed points of the $Z_2$ symmetry, it can be shown that
only the relative displacement of the two branes is physical, which
we call the radion.
There are two distinctively different kinds of displacement of
a brane: the ``bend'' and ``fluctuation''. The bend is a
type of displacement due to inhomogeneities of the
matter energy-momentum tensor on the brane. The trace of the
energy-momentum acts as an additional tension of the brane, and the
brane must ``bend'' accordingly. The relative bend is described
by the mode of radion that couples with the source on the
branes, and it can be written as a functional of the
energy-momentum tensor\cite{GT,GS}.
In the context of the quasi-localized gravity discussed by Gregory,
Rubakov \& Sibiryakov\cite{GRS}, the role of this type of radion
has been extensively studied\cite{CEH1,DGP,CEH2,PRZ,KMPP}.
On the other hand, the ``fluctuation'' is a type of displacement that
is purely geometrical, which obeys a free wave equation without source.
The relative displacement of this kind is the mode of radion
which we shall discuss in this paper.
This mode of radion was studied first by Charmousis, Gregory and Rubakov
for the RS branes whose effective radion mass is zero
by solving the field equations for the RS branes \cite{CGR} and by
Chacko and Fox\cite{CF} for the dS and AdS branes whose effective radion
mass squared are negative and positive, respectively.

The fact that the radion mass squared is negative (or zero)
suggests the (marginal) instability of the two brane system.
In fact, in the case of the original RS flat two brane
model\cite{RS1}, the negativity of the Brans-Dicke parameter on the
negative tension brane\cite{GT} can be regarded as a result of
this marginal instability.
To recover the stability, Goldberger \& Wise introduced a bulk
scalar field\cite{GW} that couples to the branes in such a way
that the distance between the two branes is stabilized.
However, it should be also noted that the effective Brans-Dicke
parameter on the positive tension brane is positive and it can
be large enough to be consistent with experiments for the separation
of the branes larger than the AdS curvature radius
at least at the linear perturbation order\cite{GT}.
Hence a stabilization mechanism may be unnecessary if we live on
the positive tension brane.
We therefore do not introduce a stabilization mechanism.

We shall work on the system that consists of an $(n+1)$-dimensional
AdS bulk spacetime bounded by two branes of constant curvature
that are fixed points of the $Z_2$-symmetry.
The zero-curvature branes
correspond to the flat RS branes\cite{RS1,RS2},
while the positive-constant curvature branes correspond
to the dS branes\cite{GaS}.
Our main concern is of course the dS brane case, but
we treat the flat brane case simultaneously to make clear
the similarities and differences between the two cases.

The dS brane case is of particular interest because it
gives a good model of braneworld inflation. In the
standard 4-dimensional inflation, the quantum vacuum
fluctuations play a very important role. It is therefore
natural to ask if the quantum radion fluctuations play
an important role, if not disastrous, in the braneworld
inflation.

It should be mentioned that historically a very similar situation
was analyzed by Garriga and Vilenkin \cite{GV91,GV92} in which
they considered the fluctuations of
a thin domain wall in $(N+1)$ spacetime dimensions. Although they
assumed the Minkowski background, many of the results
obtained there apply equally to the present case.
In particular, they showed that the wall fluctuation mode
is represented by a scalar field living on the $N$-dimensional
de Sitter space which describes the internal metric on the
domain wall, and the scalar field has the negative mass squared
$-NH^2$.

The paper is organized as follows.
In Sec.~\ref{sec:bg}, we describe the background spacetime
and our notation. In Sec.~\ref{sec:pert}, we solve the perturbation
equation in the bulk that describes the radion mode.
We find there is a gauge degree of freedom that should be carefully
treated in the dS brane case in contrast to the flat brane case
where no such subtlety arises.
In Sec.~\ref{sec:qradion}, assuming the
dependence on the extra dimensional coordinate that solves the
perturbation equation in the bulk, we derive the effective action
for the radion and quantize it. In Sec.~\ref{sec:omega},
based on the result obtained in Sec.~\ref{sec:qradion}, we
evaluate the effective energy density of the radion on the brane
which is present at the linear order in the radion amplitude, and
estimate its effect by calculating the rms value. We find
the effect remains small for reasonable values of the model
parameters even though the radion itself is unstable.
In Sec.~\ref{sec:summary}, we summarize our results and
discuss the implications.

\section{Background}\label{sec:bg}

First, we summarize the basic equations of the system and explain our
notation. The system we consider is a $Z_2$ symmetric
$(n+1)$-dimensional bulk spacetime
with two $(n-1)$-dimensional branes as the
fixed points of the symmetry. The bulk metric $g_{ab}$ obeys the
$(n+1)$-dimensional Einstein equation:
\begin{eqnarray}
 \label{bulkEin}
  {}^{(n+1)}G_{ab}+\Lambda_{n+1}g_{ab}=\kappa^2\,T_{ab}\,,
\end{eqnarray}
where $\kappa^2$ is the $(n+1)$-dimensional gravitational constant,
$\Lambda_{n+1}$ is the $(n+1)$-dimensional cosmological constant which
we assume to be negative, and $T_{ab}$ is localized on the branes.
We use the Latin indices for $(n+1)$-dimensional tensor fields in the bulk
and the Greek indices for $n$-dimensional tensor fields on the brane.
Denoting the $n$-dimensional metric on the brane by $q_{\mu\nu}$,
we decompose the localized energy momentum tensor $T_{ab}$ into
the tension part $-\sigma q_{\mu\nu}$ and the matter part
$\tau_{\mu\nu}$. Then Eq.~(\ref{bulkEin}) reduces to
\begin{eqnarray}
 \label{branEin}
{}^{(n)}G_{\mu\nu}+\Lambda_{n} q_{\mu\nu}
=\kappa_{n}^2\tau_{\mu\nu}+\kappa^4\pi_{\mu\nu}-E_{\mu\nu}
\end{eqnarray}
on the brane\cite{SMS}, where $\pi_{\mu\nu}$ is a tensor field quadratic in
$\tau_{\mu\nu}$, and $E_{\mu\nu}$ is the projected $(n+1)$-dimensional Weyl
tensor defined by $E_{\mu\nu}:={}^{(n+1)}C^a_{\ \mu b\nu}n_{a}n^{b}$ with
$n^a$ being the unit vector normal to the brane. The constants
$\Lambda_n$ and $\kappa_n^2$ are related to the basic constants of the
systems as\footnote{There is a typo in the corresponding equation in
\cite{GS}.}
\begin{eqnarray}
 \label{branLamb}
  \Lambda_n := \frac{n-2}{n}\kappa^2
     \left( \Lambda_{n+1}+\frac{n}{8(n-1)}\kappa^2\sigma^2\right),\ \ \
   \kappa_n^2 := \frac{n-2}{4(n-1)}\sigma \kappa^4.
\end{eqnarray}
It may be noted that although the decomposition of $T_{ab}$ is not unique,
there is no arbitrariness in the metric on the brane.

We consider branes with $\tau_{\mu\nu}=0$.
Looking at Eq.~(\ref{branEin}), we see that the $E_{\mu\nu}$ term can
be considered as the effective energy-momentum tensor on the brane
induced by the bulk gravitational field,
\begin{eqnarray}
   \tau^E_{\mu\nu}:=-\frac{1}{\kappa_n^2}E_{\mu\nu}\,.
\label{effTau}
\end{eqnarray}
The background we consider is the RS brane system and the dS brane system.
To be more precise, the bulk is the
$(n+1)$-dimensional anti-de~Sitter spacetime (AdS$_{n+1}$) whose metric
may be written as
\begin{eqnarray}
 && ds^2=\tilde{g}_{ab}dx^a dx^b
        =b^2(z)\left[dz^2 + \gamma_{\mu\nu} dx^\mu dx^\nu\right]\,;
    \qquad b(z)=\frac{\ell\sqrt{\cal K}}{\sinh \sqrt{\cal K}z}\,,
    \quad  \ell=\left[-\frac{n(n-1)}{2\Lambda_{n+1}}\right]^{1/2}
\end{eqnarray}
where $\gamma_{\mu\nu}$ is the metric of a Lorentzian $n-$dimensional
constant curvature space with curvature ${\cal K}=0$ or ${\cal K}=1$,
and $\ell$ is the curvature radius of AdS$_{n+1}$.
The function $b(z)$ is called the warp factor.
The two branes are placed at the coordinates $z=z_+$ and $z=z_-$
($z_+<z_-$) with their tensions and cosmological constants given by
\begin{eqnarray}
  \sigma_{\pm}=\pm\frac{2(n-1)}{\kappa^2\ell}\cosh(\sqrt{\cal K}z_{\pm})
  \qquad\mbox{and}\qquad
\Lambda_{n,\pm}=\frac12{(n-1)(n-2){\cal K}\over b^2(z_{\pm})}\,,
\end{eqnarray}
which satisfy the effective Einstein equations with $\tau_{\mu\nu,\pm}=0$
and $E_{\mu\nu,\pm}=0$. The choice ${\cal K}=0$ corresponds to the RS
brane system while ${\cal K}=1$ corresponds to the dS brane system.
Note that the $n$-dimensional gravitational constant $\kappa_n^2$ on
the positive tension dS brane is related to that on
the positive tension RS brane, say $\kappa_{n,\rm{RS}}^2$,
as $\kappa_n^2=\kappa_{n,{\rm RS}}^2\cosh z_+$.

Although we perform most of our calculation in the coordinate system
with $z$, sometimes it is more convenient to use the coordinate system
with the proper distance coordinate $r$ defined by
\begin{eqnarray}
     dr=-b(z)dz\,.
\end{eqnarray}
It is also useful to introduce the rate of
the change in the warp factor $b(z)$ in the $r$ coordinate,
which we shall denote by $J(z)$,
\begin{eqnarray}
   J(z):=\frac{\ppp_r b}{b}=-\frac{\ppp_z b}{b^2}
=\frac{\cosh\sqrt{\cal K}z}{\ell}\,.
\end{eqnarray}

\section{Radion mode in the bulk}\label{sec:pert}

In this section, we consider the gravitational perturbation on the
background described in the previous section.
Our analysis is a generalization and a reformulation of \cite{CGR}
which discussed the RS brane system $({\cal K}=0)$.
The 4-dimensional dS and AdS brane system was discussed in \cite{CF}.

We denote the gravitational perturbation by $h_{ab}$ to the background
metric $\tilde{g}_{ab}$, i.e., $g_{ab}=\tilde{g}_{ab}+h_{ab}$ where
$g_{ab}$ is the full metric. We choose the RS gauge with respect to the
positive or negative tension branes and denote the gravitational
perturbation in this gauge by $h_{\mu\nu}^{[\pm]}$
 and the coordinates by $\{r_{[\pm]},x^\mu_{[\pm]}\}$.
The RS gauge condition is
\begin{eqnarray}
  h_{55}^{[\pm]}=h_{5\mu}^{[\pm]}=
  h^{[\pm]\mu}_{\ \ \ \ \mu}=h^{[\pm]\,;\nu}_{\mu\nu}=0\,.
\label{RSgaugeDEF}
\end{eqnarray}
In the RS gauge, the Einstein equations~(\ref{bulkEin})
simplify to
\begin{eqnarray}
&& \left[ \frac{1}{b^{n-1}}b\ppp_r b^{n-1}b\ppp_r+\Box_n -2{\cal K}\right]
         \frac{h_{\mu\nu}^{[\pm]}}{b^2}=0\,.
\label{EOMhmnRS}
\end{eqnarray}
Since $h_{ab}^{[\pm]}n^a=0$, we may use Eq.~(A23)
of \cite{GS} to obtain the boundary condition on $h_{ab}^{[\pm]}$.
Expressing the perturbed locations of the branes
in terms of $\varphi^{[\pm]}(x)$ as
\begin{eqnarray}
  r_{[+]}=r_{+}+\varphi^{[+]}(x)
     \quad\mbox{and}\quad
  r_{[-]}=r_{-}+\varphi^{[-]}(x)\,,
\end{eqnarray}
where $r_{\pm}$ denote the positions of the background branes,
we have
\begin{eqnarray}
  b\ppp_r\left[\frac{ h_{\mu\nu}^{[\pm]}}{b^2}\right]=
       \frac{2}{b}\varphiMuNu{[\pm]}
  \qquad \mbox{at}\quad r_{[\pm]}=r_{\pm}\mp0\,,
 \label{JuncCond}
\end{eqnarray}
where
\begin{eqnarray}
{\cal L}_{\mu\nu}:=D_\mu D_\nu+{\cal K}\gamma_{\mu\nu}\,.
\end{eqnarray}
Note that $\tau_{\mu\nu,\pm}$ is set to zero as we are interested in
the gravitational perturbation that does not couple with the matter
on the brane.
It should be noted here that Eq.~(A23) of \cite{GS} is defined from
the side with the smaller value of $r$, while in this paper
the boundary condition at the negative tension brane is
given from the larger value of $r$; $r_{[-]}=r_{-}+0$, hence
the signature is reversed.
In passing, we also note that Eq.~(\ref{JuncCond}) may
be derived by first introducing the Gaussian normal coordinates
with respect to the branes and transforming them to the RS gauge,
as was done in \cite{CGR}.

With the junction condition (\ref{JuncCond}) at hand, we consider a
particular solution of the form,
\begin{eqnarray}
 h^{[\pm]}_{\mu\nu}
      =b^{2}u(z)\varphiMuNu{[\pm]}(x^\mu)\,.
\label{FORMhmn}
\end{eqnarray}
The traceless condition on the metric
perturbation in the RS gauge demands $\varphi^{[\pm]}$ to satisfy
\begin{eqnarray}
  [-\Box_n-n{\cal K}]\varphi^{[\pm]}=0\,,
\end{eqnarray}
which implies
\begin{eqnarray}
 [-\Box_n+n{\cal K}]\varphiMuNu{[\pm]}=0\,.
  \label{varphiDEF}
\end{eqnarray}
Then Eq.~(\ref{EOMhmnRS}) reduces to
\begin{eqnarray}
  \left[ \frac{1}{b^{n-1}}\ppp_z b^{n-1}\ppp_z+(n-2){\cal K}\right]
       u(z)=0\,.
\end{eqnarray}
The general solution to this equation is obtained as follows.
First, it is easy to see that
$u^{(1)}(z)=\ell J(z)$ is a solution.
Note here the identity $\ppp_z J(z)={\cal K}b^{-1}(z)$. Then the
other independent solution, $u^{(2)}(z)$,
may be found using the Wronskian of the above equation, which is
\renewcommand{\arraystretch}{2}
\begin{eqnarray}
  u^{(2)}_n(z)
    &&=\ell^{\,n-2}J(z)\int^z dz'\, J^{-2}(z')b^{1-n}(z')\,.
\label{psi(2)def}
\end{eqnarray}
Thus, the general solution of the form (\ref{FORMhmn}) is
\begin{eqnarray}
  h^{[\pm]}_{\mu\nu}= b^2 \ell^{-1}\left(C^{[\pm]} u^{(1)}(z)
                  + D^{[\pm]} u^{(2)}_n(z)\right)\varphiMuNu{[\pm]}\,,
\label{h+-mn}
\end{eqnarray}
where $C^{[\pm]}$ and $D^{[\pm]}$ are constants.

Note that there is an ambiguity in $u^{(2)}(z)$ which depends on
the choice of the integration constant. The change of it induces
the change of the coefficient $C^{[\pm]}$. This implies the solution
$u^{(1)}$ represents a gauge degree of freedom.
In fact, if we evaluate $E^{[\pm]}_{\mu\nu}$, we obtain
\begin{eqnarray}
  E^{[\pm]}_{\mu\nu}=-\frac1{n-1}[\ppp_z^2-{\cal K}]
                          \left(\frac{h^{[\pm]}_{\mu\nu}}{b^2}\right)
                    =-(b/\ell)^{2-n}\ell^{-1}D^{[\pm]}\varphiMuNu{[\pm]}\,,
\label{E+-mn}
\end{eqnarray}
which is independent of $C^{[\pm]}$.
Note also that the $z$-dependence of $E_{\mu\nu}$ is clearly consistent
with the boundary condition at the branes with no source on it,
$\ppp_z( b^{n-2}E_{\mu\nu})=0$\cite{GS}.

Since $E_{\mu\nu}$ is a gauge-invariant tensor field,
we see that the displacement of the branes should be related as
\begin{eqnarray}
  D^{[+]}\varphi^{[+]}=D^{[-]}\varphi^{[-]}\,.
\label{Dvp=Dvp}
\end{eqnarray}
With the above relation at hand, we can use a
residual gauge degree of freedom in the RS gauge
to bring the two different sets of the
coordinates $\{x^a_{[\pm]}\}$ to the same coordinates, say $\{x^a\}$.
This is done by an infinitesimal coordinate transformation,
\begin{eqnarray}
&&
  r\rightarrow r+\hat{\xi}^r(x^\mu)\,,\qquad
  x^{\mu}\rightarrow x^{\mu}
            +\frac{\ell}{2}\,{\cal B}^{-2}\hat{\xi}^{r|\mu}(x^\mu)
            +\hat{\xi}^{\mu}(x^\mu)\,;
\label{habna=0}\\
&&\hspace{15ex}
  \hat{\xi}^r=\frac{\cal K}{2}\left[C^{[\pm]}-\alpha D^{[\pm]}\right]
                  \varphi^{[\pm]}\,,\qquad
  \hat{\xi}_{\mu}=\frac{1}{2\ell}
         \left[C^{[\pm]}-\alpha D^{[\pm]}\right]\varphi^{[\pm]}_{|\mu}\,,
\end{eqnarray}
where $\alpha$ is the yet remaining gauge parameter and
the function ${\cal B}(z)$ is defined as
\renewcommand{\arraystretch}{1}
\begin{eqnarray}
 {\cal B}^{-2}:=-\frac{\ell}{2}\left[\int_0^z b^{-2}dr\right]^{-1}
           =\frac{2}{\ell^2}\frac{\cosh\sqrt{\cal K}z-1}{\cal K}
      =\left\{
         \begin{array}{cl}
           \displaystyle
              b^{-2}                       & ({\cal K}=0)\\
           \displaystyle
             \frac2{\ell}(J-\ell^{-1}) & ({\cal K}=1)\,.
         \end{array}
      \right.
\end{eqnarray}
The metric perturbation in the unified coordinates $\{x^a\}$ is
\begin{eqnarray}
   h_{\mu\nu}= b^2 \ell^{-1}\left(\alpha u^{(1)}(z)
                +  u^{(2)}(z)\right)D^{[\pm]}\varphiMuNu{[\pm]}\,,
\end{eqnarray}
and the branes are now placed at
\begin{eqnarray}
%    \left(\begin{array}{l}
%      \mbox{positive}\\
%      \mbox{negative}
%    \end{array}\right)
%     \mbox{ brane at }
  r=r_{\pm}
     +\left(1+\frac{\cal K}{2}\left[C^{[\pm]}-\alpha 
D^{[\pm]}\right]\right)
                 \varphi^{[\pm]}\,.
\end{eqnarray}
We note that the positions of the two branes cannot fluctuate independently
in this unified gauge because of Eq.~(\ref{Dvp=Dvp}).
This shows the nature of the radion that it describes the relative
displacement of the branes.

With the above form of the metric perturbation and the locations of the
branes, we find that the boundary condition on $h_{ab}$ gives
\begin{eqnarray}
  D^{[\pm]}=-\frac{2\ell}
           {\alpha{\cal K}\ell+b_\pm\ppp_z u^{(2)}(z_\pm)}
  \left(1+\frac{\cal K}{2}\left[C^{[\pm]}-\alpha D^{[\pm]}\right]\right)\,,
\label{D+-}
\end{eqnarray}
which implies
\begin{eqnarray}
1+{{\cal K}\over2}C^{[\pm]}
+{b_{\pm}\partial_z u^{(2)}(z_{\pm})\over2\ell}D^{[\pm]}=0\,.
\label{CDrel}
\end{eqnarray}
An apparent complication in the case of ${\cal K}=1$ is that
this seems to constrain the values of $C^{[\pm]}$ which
is just unphysical, gauge parameters. This can be resolved by considering
the
integration constant term in $u^{(2)}$ defined in Eq.~(\ref{psi(2)def}).
For a change of $C^{[\pm]}$ as
\begin{eqnarray}
  C^{[\pm]}\to C^{[\pm]}+\beta\,,
\end{eqnarray}
we consider the simultaneous change
of the integration constant term in $u^{(2)}$ as
\begin{eqnarray}
 u^{(2)}\to u^{(2)}-(D^{[\pm]}\beta)\,u^{(1)}\,,
\end{eqnarray}
which results in
\begin{eqnarray}
\partial_z u^{(2)}\to \partial_z u^{(2)}-D^{[\pm]}\beta
                              {{\cal K}\ell\over b}\,.
\label{dzpchange}
\end{eqnarray}
This combined change of $C^{[\pm]}$ and the integration constant term in
$u^{(2)}(z)$ leaves Eq.~(\ref{CDrel}) as well as
$h^{[\pm]}_{\mu\nu}$ unchanged.
One can also see from Eq.~(\ref{dzpchange}) the reason for the
simplification in the case of ${\cal K}=0$, i.e., the change of
$u^{(2)}$ does not affect $\partial_z u^{(2)}$
if ${\cal K}=0$.

Introducing $\tilde\varphi(x)$ defined as
\begin{eqnarray}
    \tilde\varphi
        :=-{D^{[+]}\varphi^{[+]}\over2\ell}
=-{D^{[-]}\varphi^{[-]}\over2\ell}\,,
\end{eqnarray}
the results are summarized as
\renewcommand{\arraystretch}{2}
\begin{eqnarray}
&&
 \left\{
 \begin{array}{l}
   \bullet\ \
      h_{\mu\nu}=- 2 b^2(\alpha u^{(1)}+u^{(2)} )
                        {\cal L}_{\mu\nu}\tilde{\varphi}
          \quad\mbox{and}\quad
      E_{\mu\nu}=2 (b/\ell)^{2-n}{\cal L}_{\mu\nu}\tilde{\varphi}\\
   \bullet\
      \renewcommand{\arraystretch}{1}
      \left(\begin{array}{l}
              \mbox{positive}\\
              \mbox{negative}
            \end{array}\right)
         \mbox{ brane at } r=r_{\pm}+L(z_{\pm};\alpha)\tilde\varphi
 \end{array}\right.
\label{result}
\end{eqnarray}
where $L(z;\alpha):=\alpha\ell{\cal K}+b(z)\ppp_z u^{(2)}(z)$
and $\tilde\varphi$ satisfies
\begin{eqnarray}
\left[-\Box_n-n{\cal K}\right]\~\varphi(x^\mu)=0\,.
\label{EOMvarphi}
\end{eqnarray}
As it should be clear by now, the parameter $\alpha$ describes
the residual freedom of the RS gauge in the unified coordinates.
corresponding to the coordinate transformation,
$r\rightarrow r+{\cal K}f(x^\mu)$ and
$x^{\mu}\rightarrow x^{\mu}+ Jf^{|\mu}(x^\mu)$
with $f(x^\mu)$ satisfying
${\cal L}^{\mu}_{\ \mu}f(x^\mu)=0$.

\section{Quantum Radion}\label{sec:qradion}

Given the initial data for $\~\varphi$, we obtain the full time
evolution of the fluctuations of the brane by solving
Eq.~(\ref{EOMvarphi}). As the initial data,
it is natural to assume that $\varphi$ is classically null,
and only the quantum vacuum fluctuations are present.

To consider the quantum fluctuations of the brane, we need the
action for $\~\varphi$ in the first place.
We obtain it by substituting the classical solution (\ref{result})
into the original action of the system, but
without constraining $\~\varphi(x)$ to satisfy Eq.~(\ref{EOMvarphi}).
This is because thus obtained action is already maximized except for the
degree of freedom for the brane to fluctuate and hence is adequate for
the quantization of such a degree of freedom\cite{RMW}.

Our system consists of identical bulk spacetimes ${\cal M}_1$ and ${\cal 
M}_2$
with branes $\Sigma_+$ and $\Sigma_-$ that are the boundary hypersurfaces
of ${\cal M}_1$ and ${\cal M}_2$ with
tensions of $\sigma_+$ and $\sigma_-$, respectively;
$\ppp{\cal M}_1=\Sigma_+\cup\Sigma_-$ and 
$\ppp{\cal M}_2=\Sigma_+\cup\Sigma_-$. 
The action of such
a system is given by
\begin{eqnarray}
  && \hspace{15ex}
       I = I_{R\mbox{-}2\Lambda,1}+I_{R\mbox{-}2\Lambda,2}
         + I_{K,1}-I_{K,2}+I_{\sigma+}+I_{\sigma-}\,;\nnb\\
  &&
     I_{R\mbox{-}2\Lambda,1}
          :=\frac{1}{2\kappa^2}\int_{{\cal M}_{1}}
                \hspace{-2.5ex}\sqrt{-g}\, dr\, d^n\! x\,
              \Big[{}^{(n+1)}R-2\Lambda_{n+1}\Big]\,,
      \quad
     I_{K,1}:=\frac{1}{\kappa^2}\int_{\ppp{\cal M}_1}
                \hspace{-3.5ex}\sqrt{-q}\,d^n\!x\, K\,,
      \quad
     I_{\sigma\pm}
          :=-\int_{\Sigma_\pm}\hspace{-2.5ex}\sqrt{-q}\,
          d^n\!x\, \sigma_{\pm}\,,
\label{basicAction}
\end{eqnarray}
where $q_{ab}:=g_{ab}-n_a n_b$,
$K_{ab}:=q_a^{\ c}q_b^{\ d}\V_c n_d$ and $n^a$ is a unit
vector normal to the branes, pointing from
${\cal M}_1$ to ${\cal M}_2$.
The substitution of $h_{\mu\nu}$ and the locations of the branes
given by Eq.~(\ref{result}) into the action (\ref{basicAction}),
We find
\begin{eqnarray}
   I^{(2)}[\~\varphi]
     &=&\frac{(n-1)(n-2)\ell^{n-1}}{\kappa^2}N_{(n-2){\cal K}}
        \int d^n\!x \sqrt{-\gamma}\,
               \~\varphi\bigl[\Box_n+n{\cal K}\bigr]\~\varphi\,,
\label{pre-action}
\end{eqnarray}
where
\begin{eqnarray}
   N_{(n-2){\cal K}}
      :=\ell^{n-3}\int^{z_-}_{z_+}dz' b^{3-n}
       =\frac{1}{(n-2)\ell}
          \Big[b_-\ppp_z u^{(2)}(z_-)-b_+\ppp_z u^{(2)}(z_+)\Big].
\end{eqnarray}
The normalization factor $N_{(n-2){\cal K}}$ is found to be
identical to the one introduced in \cite{GS} for the normalization of the
radion mode of $E_{\mu\nu}$. Details of the derivation and its
comparison with those by Chiba\cite{CH} and by Chacko and Fox\cite{CF}
are given in Appendix~\ref{CalAction}.

Since $N_{(n-2){\cal K}}>0$, we may normalize $\~\varphi$ as
\begin{eqnarray}
   \psi(x^\mu):=
    \sqrt{{2(n-1)(n-2)N_{(n-2){\cal K}}\,\ell\over\kappa^2}
     \left({\ell\over b_+}\right)^{n-2}}\,\~\varphi(x^\mu),
\end{eqnarray}
such that the action reduces to that of a scalar field
$\psi(x)$ with the mass squared
$m^2:=-n{\cal K}b_+^{-2}$ on the positive tension brane:
\begin{eqnarray}
    I^{(2)}[\psi]
   &=&
      \int d^n\! x \sqrt{-q_+}\,
          \bigg[
-\frac12q_+^{\mu\nu}\ppp_\mu\psi\ppp_\nu\psi-\frac12m^2\psi^2
           \bigg]\,,\label{ActionPsi}
\end{eqnarray}
where $q_{+\mu\nu}:=b^2_+\gamma_{\mu\nu}$ is the intrinsic metric of the
positive tension brane.
Here and below, we focus on the positive tension brane, but the
parallel arguments apply to the negative tension brane except for
the difference in the normalization factor.

In terms of $\psi(x^\mu)$, we see from Eq.~(\ref{result}) that
$E_{\mu\nu}$ is given by
\begin{eqnarray}
   E_{\mu\nu}=N_E\,\left({b_+\over b}\right)^{n-2}
{\cal L}^{\mathrm{TL}}_{\mu\nu}\,\psi(x)\,,
\label{EmnPhi}
\end{eqnarray}
where
\begin{eqnarray}
&&N_E=\sqrt{{2\over(n-1)(n-2)N_{(n-2){\cal K}}}{\kappa^2\over\ell}
\left({\ell\over b_+}\right)^{n-2}}\,,
\nonumber\\
&&{\cal L}^{\mathrm{TL}}_{\mu\nu}={\cal L}_{\mu\nu}
-{\gamma_{\mu\nu}\over n}\gamma^{\alpha\beta}{\cal L}_{\alpha\beta}
=D_\mu D_\nu -{\gamma_{\mu\nu}\over n}\Box_n\,.
\end{eqnarray}
Here it may be appropriate to make a comment on the form of
Eq.~(\ref{EmnPhi}) in which the operator ${\cal L}_{\mu\nu}$
is replaced by its traceless part.
In the previous section,
we considered only the classical solution for the radion fluctuations,
hence we were allowed to use the equation of motion to express $E_{\mu\nu}$
in terms of the radion field, and the equation of motion for the radion was
derived from the traceless condition of the RS gauge.
When we consider the quantum radion, however,
$E_{\mu\nu}$ must be traceless at the off shell level as well.
This means that the traceless condition of the RS gauge
should not be used to express $E_{\mu\nu}$ in terms of the radion field.
Instead, one has to go back to the original definition of the Weyl tensor
in terms of the metric perturbation whose trace may not be null.
 This is the reason for the
appearance of the traceless part of ${\cal L}$ in Eq.~(\ref{EmnPhi}).

Now that we have obtained the action for the fluctuation of the brane,
which is of the form of the scalar field action,
we may canonically quantize it in the ordinary way.
Adopting the flat coordinate chart on the brane, which is specified as
\begin{eqnarray}
  ds_{+}^2= b_+^2\gamma_{\mu\nu}dx^\mu dx^\nu
            = a^2(\eta)\left[ -d\eta^2+(d\bm{x})^2\right]
  \qquad
    \left( a(\eta):=\frac{b_+}{(1-{\cal K})-{\cal K}\eta}\,,
   \quad   H_+:=\frac{\ppp_\eta a}{a^2}={\cal K}b_+^{-1}\right)\,,
\end{eqnarray}
we obtain
\begin{eqnarray}
   \widehat{\psi}(\eta,\bm{x})=\int d^{n-1}\bm{k}
       \left[ \hat{a}_{\bm{k}}\psi_{k}(\eta)e^{i\bm{k}\cdot\bm{x}}
         +\hat{a}_{\bm{k}}^\dagger\psi_{k}^* (\eta)e^{ -i\bm{k}\cdot\bm{x} 
}
\right],
\end{eqnarray}
where $\hat{a}_{\bm{k}}$ and $\hat{a}_{\bm{k}}^\dagger$ satisfy
the commutation relations,
\begin{eqnarray}
   [\hat{a}_{\bm{k}},\hat{a}_{\bm{k}'}^\dagger]=\delta(\bm{k}-\bm{k}')\,,
\quad [\hat{a}_{\bm{k}},\hat{a}_{\bm{k}'}]
=[\hat{a}_{\bm{k}}^\dagger,\hat{a}_{\bm{k}'}^\dagger]=0\,,
\end{eqnarray}
and $\{\psi_{k}(\eta)\}$ satisfies the equation of motion,
\begin{eqnarray}
 &&\left[ \frac{d^2}{d\eta^2}-\frac{(n-2){\cal K}}{\eta}\frac{d}{d\eta}
               +k^2-\frac{n{\cal K}}{\eta^2}\right]\psi_{k}(\eta)=0\,,
\label{psikeom}
\end{eqnarray}
with the normalization condition,
\begin{eqnarray}
\psi_{k}{d\over d\eta}\psi_{k}^*-\psi_{k}^*{d\over d\eta}\psi_{k}
={2i\over a^{n-2}}\,.
\end{eqnarray}
As a natural choice for the vacuum
we choose the Bunch-Davis vacuum, for which we have
\renewcommand{\arraystretch}{2.3}
\begin{eqnarray}
   \psi_k(\eta)
    =\left\{
       \begin{array}{ll}
          \displaystyle
          \frac{e^{-ik\eta}}{\sqrt{2k}} & ({\cal K}=0),\\
         \displaystyle
          \frac{\sqrt{\pi}}{2}b_+^{-(n-2)/2}(-\eta)^{(n-1)/2}
               H_{(n+1)/2}^{(1)}(-k\eta)\quad & ({\cal K}=1),
       \end{array}
     \right.
\label{BDvac}
\end{eqnarray}
where $H_{\lambda}^{(1)}(z)$ is the Hankel function of the first kind.
For ${\cal K}=1$, this would correspond to choosing
the de~Sitter invariant vacuum if the mass squared were positive.
Note that the rms amplitude of the vacuum fluctuations per logarithmic
interval of $k$ exhibits an infrared divergence in the limit $\eta\to-0$,
\begin{eqnarray}
\sqrt{k^{n-1}|\psi_k(\eta)|^2}\propto {1\over |k\eta|}\,.
\label{radinst}
\end{eqnarray}

We now focus on the case ${\cal K}=1$, which is of cosmological interest,
and consider the effect of the quantum radion on the brane.
As noted, $\psi$ shows an instability which
grows as $(-\eta)^{-1}\propto a(\eta)$.
However, what one can detect on the brane is
$E_{\mu\nu}$ given by Eq.~(\ref{EmnPhi}) which acts as the
effective energy-momentum tensor, Eq.~(\ref{effTau}).
To see its behavior, let us spell out the components of
$E_{\mu\nu}$ explicitly.
The components for each $k$ mode are
\begin{eqnarray}
E_{00}
&&=N_E\left[{n-1\over\eta}\left({\partial\over\partial\eta}
+{1\over\eta}\right)-k^2\right]
\psi_kY\,,
\nonumber\\
E_{0j}
&&=N_E(-k)\left[{\partial\over\partial\eta}+{1\over\eta}\right]
\psi_kY_{j}\,,
\nonumber\\
E_{ij}
&&=N_Ek^2\psi_kY_{ij}
+N_E\delta_{ij}\left[{1\over\eta}\left({\partial\over\partial\eta}
+{1\over\eta}\right)-{k^2\over n-1}\right]\psi_kY
\label{Emnform}
\end{eqnarray}
where Eq.~(\ref{psikeom}) has been used, and $Y$, $Y_j$, and $Y_{ij}$
stand for
\begin{eqnarray}
Y=e^{i\bm{k}\cdot\bm{x}}\,, \quad
Y_j=-i{k_j\over k}Y\,,\quad
Y_{ij}=\left(-{k_ik_j\over k^2}+{\delta_{ij}\over n-1}\right)Y\,.
\end{eqnarray}
Inserting the explicit form of $\psi_k$ given in Eq.~(\ref{BDvac}),
we find
\begin{eqnarray}
E_{00}
&&=\tilde N_E\,k^2(-k\eta)^{(n-1)/2}
H^{(1)}_{(n-3)/2}(-k\eta)\,Y\,,
\quad
E^{L}_{ij}={\delta_{ij}\over n-1}E_{00}\,,
\nnb\\
E_{0j}
&&=\tilde N_E\,k^2(-k\eta)^{(n-1)/2}
H^{(1)}_{(n-1)/2}(-k\eta)\,Y_j\,,
\nnb\\
E^{T}_{ij}
&&=\tilde N_E\,k^2(-k\eta)^{(n-1)/2}
H^{(1)}_{(n+1)/2}(-k\eta)\,Y_{ij}\,,
\label{Emnexp}
\end{eqnarray}
where $E^{L}_{ij}$ and $E^{T}_{ij}$ denote the trace and traceless part
of $E_{ij}$, respectively, and
\begin{eqnarray}
\tilde N_E=N_E{\sqrt{\pi}H_+^{(n-2)/2}\over2k^{(n-1)/2}}\,.
\end{eqnarray}
{}From the above, and noting $H^{(1)}_\nu(z)\sim z^{-\nu}$ for
$z\to0$, we see that $E^{T}_{ij}$, which represents
the effective anisotropic stress, dominates over the other components
for $k|\eta|\to0$, i.e., on super Horizon scales, though
its physical amplitude ($E^{T}_{ij}/a^2$) decays as $1/a$.

As for the effective energy density induced by the radion,
it decays rapidly as $a^{-3}$.
However, it may be said that
the decay rate is not fast enough in the following sense.
Since $E_{\mu\nu}$ is traceless and conserved
at the linear perturbation order, one can regard it as the
energy momentum tensor of a radiation-like fluid.
If it should behave exactly as radiation, the energy density
would decay as $a^{-n}$ instead of $a^{-3}$.
Of course the decay rate $a^{-3}$ does not mean
$E_{\mu\nu}$ behaves as a dust matter.
The fact that the decay rate is slower than
the standard radiation fluid (for $n\geq4$) is caused by a strong
anisotropic stress present in the case of radion.
Thus, in this sense, the instability
of radion indeed shows up on the brane, but the effect turns out to
be rather mild because of the nature of
de~Sitter space in which the background energy density remains constant
in time.
We defer the detailed discussions on this point to Appendix \ref{DecayEmn}.

To summarize, we conclude that the instability of the quantum radion
induces a large anisotropic stress on the brane, but
because of the rapid expansion of the brane it is not strong enough
to give rise to a gravitational instability on the brane.

\section{Cosmological effect of quantum radion}\label{sec:omega}

In the previous section, we have seen that the quantum
radion fluctuations do not cause a gravitational instability on the brane.
However, this does not necessarily mean the effect is always small.
In particular, if the rms value of the effective energy density
due to the radion field is comparable to the background vacuum energy
density of the de Sitter brane, the evolution of the brane will be
significantly affected. If we regard our background to represent
a braneworld undergoing cosmological inflation, this would imply
a significant modification of the braneworld inflationary scenario.
In this section, assuming the positive tension brane describes our
universe, we consider the cosmological constraints arising from
the effect of the radion quantum fluctuations.

To quantify the effect, we consider the power spectrum of the
effective energy density,
\begin{eqnarray}
  \rho_E:=\frac{1}{\kappa_n^2}E^{0}{}_{0}\,.
\end{eqnarray}
Focusing on the positive tension brane, we define the power spectrum
$P_n(k)$ of $\rho_E$ normalized by the background energy density
$\rho_\Lambda=3H_+^2/\kappa_n^2$ as
\begin{eqnarray}
   \frac{\langle (\widehat{\rho}_E)^2 \rangle}{(\rho_\Lambda)^2}
      =\frac{S_{n-1}}{(2\pi)^{n-1}}\int \frac{dk}{k} k^{n-1}P_n(k)\,,
\label{power}
\end{eqnarray}
where $S_{n-1}$ is the surface area of the $(n-1)$-dimensional unit sphere;
$S_{n-1}=2\pi^{n/2}/\Gamma(n/2)$.
We find
\begin{eqnarray}
 k^{n-1} P_n(k)
&&={2\pi\kappa_n^2H_+^{n-2}(H_+\ell)^{n-2}\over9(n-1)(n-2)^2N_{(n-2){\cal 
K}}
\cosh z_+}
(-k\eta)^{n+3}|H^{(1)}_{(n-3)/2}(-k\eta)|^2
\nnb\\
&&\mathop{\longrightarrow}\limits_{(-k\eta)\to0}~
 {2\pi\kappa_n^2H_+^{n-2}(H_+\ell)^{n-2}\over9(n-1)(n-2)^2N_{(n-2){\cal K}}
\cosh z_+}
{2^{n-3}|k\eta|^6\over\displaystyle \Gamma\left(n-{1/2}\right)^2}\,.
 \label{Ett}
\end{eqnarray}

Let us introduce the density parameter $\Omega_E$ for the rms
effective energy density of the radion field as
\begin{eqnarray}
\Omega_E:=\frac{\sqrt{\langle (\widehat{\rho}_E)^2 
\rangle}}{\rho_\Lambda}\,,
\end{eqnarray}
which is obtained by integrating $k^{n-1}P_n(k)$ over $k$ and
taking the square root of it.
When $\Omega_E$ is comparable to, or larger than the order of unity,
the fluctuation of the brane is non-negligible.
If we integrate the above over $k$, we encounter the divergence from
large $k$.
However, this ultraviolet divergence is the same as the
usual one in the Minkowski background.
Hence, to discard the divergence,
we cut off the integration at the Hubble horizon scale;
$k= k_{H}=1/|\eta|$.
The result is
\begin{eqnarray}
 \Omega_E
&&=\sqrt{{S_{n-1}\over6(2\pi)^{n-1}}k^{n-1}P(k)\Big|_{k=1/|\eta|}}\,
\mathop{=}\limits_{n=4}\,
{2\over135}\sqrt{2\over\pi\cosh z_{+}(\cosh z_{-}-\cosh z_{+})}\,
(\kappa_4H_+)(H_+\ell)\,.
\label{OmegaE}
\end{eqnarray}
Here and below we fix the dimensions to $n=4$.
Note that
$H_+\ell=\sinh z_+$ is the ratio of the AdS$_5$ curvature radius
to the dS$_4$ curvature radius,
and $\kappa_4^2=\kappa^2\cosh z_{+}/\ell$.
If we introduce the $4$-dimensional
Planck scale in the single flat brane limit,
$\ell_{\rm pl}=\kappa_4/\sqrt{\cosh z_+}$,
we have the well-known relation, $\ell_{\rm pl}^2=\ell_5^3/\ell$.
Below we assume $\ell_{\rm pl}$
is the present-day Planck length in our universe.

To discuss the cosmological constraints, it is convenient to
introduce the $5$-dimensional gravitational length scale
$\ell_{5}=\kappa^{2/3}$.
In terms of $\ell_5$, Eq.~(\ref{OmegaE}) is re-expressed as
\begin{eqnarray}
   \Omega_E = {2\over135}\sqrt{2\over\pi}
{1\over\sqrt{\cosh z_{-}-\cosh z_{+}}}\,
(H_+\ell)^2\left({\ell_5\over\ell}\right)^{3/2}\,.
\end{eqnarray}
We see that the quantum radion effect will be negligible if
\begin{eqnarray}
\cosh z_{-}-\cosh z_{+}\gtrsim
(H_+\ell)^2\left({\ell_5\over\ell}\right)^{3/2}\,.
\end{eqnarray}
Given the ratios $H_+\ell$ and $\ell_5/\ell$,
this constrains the value of $H_-$ to be
larger than a critical value,
\begin{eqnarray}
\sqrt{1+(H_-\ell)^2}\gtrsim\sqrt{1+(H_+\ell)^2}
+(H_+\ell)^2\left({\ell_5\over\ell}\right)^{3/2}\,.
\label{critical}
\end{eqnarray}

Since $\ell_5$ is the fundamental scale of the 5-dimensional
theory, it is natural to assume $\ell_5\lesssim\ell$;
otherwise the whole system will be in the quantum gravitational
regime. Note that $(\ell_5/\ell)^3=(\ell_{\rm pl}/\ell)^2$.
With this assumption, let us consider
the following two limiting cases.
\begin{list}{}{}
\item[(a)] $H_+\ell\ll 1$:\\
 In this case, Eq~(\ref{critical}) is satisfied for
practically all values of $H_->H_+$ except for the range very close to 
$H_+$,
that is, the constraint becomes
\begin{eqnarray}
{H_-\over H_+}-1\gtrsim \left({\ell_5\over\ell}\right)^{3/2}\,.
\end{eqnarray}
\item[(b)] $H_+\ell\gg 1$:\\
In this case, we have
\begin{eqnarray}
{H_-\over H_+}-1\gtrsim (H_+\ell)\left({\ell_5\over\ell}\right)^{3/2}\,.
\end{eqnarray}
Using the equality $(\ell_5/\ell)^{3/2}=\ell_{\rm pl}/\ell$,
the right hand side is rewritten as
$H_+\ell_{\rm pl}$. For any reasonable scenario of inflation,
we must have $H_+\ell_{\rm pl}\ll1$. Hence
the constraint on the value of $H_-$ is very mild in this case as well.
\end{list}
Thus, with the assumptions that $H_+\ell_{\rm pl}\ll1$ and
$(\ell_5/\ell)^3=(\ell_{\rm pl}/\ell)^2\lesssim1$,
the constraints obtained in the above
two cases can be concisely expressed as
\begin{eqnarray}
\ln\left({H_{-}\over H_{+}}\right)
\gtrsim (1+H_+\ell){\ell_{\rm pl}\over\ell}\,.
\end{eqnarray}
To summarize, as far as the quantum radion fluctuations are concerned,
the negative tension brane can be very close to
the positive tension brane, even within a distance smaller than
the AdS$_5$ curvature radius.

\section{Summary and Discussion}\label{sec:summary}

In this paper, taking up the Randall-Sundrum type two-brane
scenario in an $(n+1)$-dimensional spacetime,
we have investigated the quantum fluctuations
of the relative displacement of the branes, which we called
the quantum radion. We have considered the cases of the
flat two-brane and de Sitter two-brane systems simultaneously.
Adopting the so-called Randall-Sundrum gauge,
we have first solved the linear gravitational perturbation equations
that describe the radion mode.

Then assuming the perturbation of the form with a fixed $z$-dependence
that solves the gravitational equation in the $z$-direction, where
$z$ is the extra-dimensional coordinate orthogonal to the branes,
we have derived the effective action for the radion.
With the effective action at hand, we have quantized the radion
assuming the radion state is the Bunch-Davis vacuum.

We have analyzed the effect of the quantum radion on the brane
using the effective Einstein equations derived by
Shiromizu, Maeda and
Sasaki\cite{SMS} in which the radion adds an effective
energy-momentum tensor at the linear order in the field amplitude.
Although the radion has the negative mass squared $-nH^2$
on the de Sitter brane where $H$ is the Hubble parameter,
we have found that the corresponding instability does not
show up on the brane. We have noted, however, that the anisotropic
stress induced by the radion is unusually large, though it still
decays in time as $a^{-1}$ for a fixed $k$, where $a$ is the cosmic
scale factor and $k$ is the comoving wavenumber,
and the rms value of the
effective energy density for a fixed $k$
decays as $a^{-3}$ irrespective of the spacetime dimensions $n$.

Focusing on the positive tension de Sitter brane with
the Hubble rate $H_+$, which models
the braneworld inflation, we have estimated the rms total energy
density of the quantum radion by integrating over
$k$ up to $aH_+$.
We have introduced the density parameter $\Omega_E$
which describes the relative magnitude of the radion energy density
to the background energy density, and discussed the condition
$\Omega_E\ll1$ on the model parameters.
For $H_+\ell_{\rm pl}\ll1$ and $\ell_5\lesssim \ell$,
which are reasonable to assume for the background to be not
in the quantum regime, we have found that
practically any choice of the location of the negative tension
brane is allowed. This implies the quantum radion does not
seriously affect the braneworld inflation scenario, at least
at the linear perturbation order.

The most intriguing question remained now is 
if the analysis here at the order of the linear
perturbation level is sufficient.
Naively, one may regard the effective action of the radion
we have obtained as a piece to be added to the total
effective action for an effective 4-dimensional theory
that includes the gravity.
Then the variation of the radion effective action with respect to the
4-metric will give the energy-momentum tensor of a scalar
field with a negative mass squared, which would grow and diverge in
contrast to the decaying effective energy-momentum tensor we analysed in
this paper. However, such an energy-momentum tensor is quadratic in the
radion, which is beyond the accuracy of our analysis at the linear-order
perturbation level.  We do need to investigate the
higher order perturbation to truly see the effect of the quantum
radion on the brane. This issue is left for future study.

\section*{Acknowledgements}
This work is supported in part by the Monbukagakusho Grant-in-Aid for
Scientific Research No.~12640269 and by the
Yamada Science Foundation.
UG is supported by the JSPS Research Fellowships for
Young Scientists, No.~2997.
We would like to thank T. Chiba, J. Garriga, K. Koyama, T. Shiromizu
and T. Tanaka for discussions.

\appendix

\section{Effective Action of radion derived and compared}\label{CalAction}

In this appendix, we give an outline of the derivation of the action
given in Eq.~(\ref{pre-action}) and compare it with the effective
actions of radion derived in \cite{CH,CF}.
The action is obtained by substituting
the results in Eq.~(\ref{result}) into the action of the system given in
Eq.~(\ref{basicAction}). Note that $\gamma_{\mu\nu}$
denotes the metric of the $n$-dimensional space-time with a constant
curvature ${\cal K}$ and $D_\mu$ the covariant derivative
with respect to $\gamma_{\mu\nu}$.
We do not demand $\~\varphi(x^\mu)$  to satisfy
the field equation~(\ref{EOMvarphi}),
so that the degree of freedom associated with the brane fluctuation
remains.

We first substitute the results given in Eq.~(\ref{result})
into the bulk actions $I_{R-2\Lambda,1}$ and $I_{R-2\Lambda,2}$ of
Eq.~(\ref{basicAction}). Carrying out the calculation with
the help of the commutation rule,
\begin{eqnarray}
\left[D_{\mu}D_{\nu}-D_{\nu}D_{\mu}\right]\omega_\rho
  =2{\cal K}\gamma_{\mu[\rho}\gamma_{\nu]\sigma}\omega^\sigma\,,
\label{commutation}
\end{eqnarray}
results in vanishing of all the second-order terms
except for the terms that can be cast into the surface term
on $\Sigma_{\pm}$.
Thus, to the second order in $\tilde\varphi$,
the action~(\ref{basicAction}) is found to have
support only on $\Sigma_{\pm}$:
\begin{eqnarray}
  && I^{(2)}=I^{(2)}_{\Sigma_+}+I^{(2)}_{\Sigma_-}\,;\nnb\\
  && \quad
      I^{(2)}_{\Sigma_\pm}
       =\frac{1}{\kappa^2}\int_{\Sigma_\pm}\hspace{-1ex} d^n x \sqrt{-q}
          \bigg[ - \frac12 \hat{k}^{\mu\nu}h_{\mu\nu}
                 + \frac12 \hat{k}_{\sigma}^{\ \sigma}h
                 + h_{\mu\nu}\varphi^{|\mu\nu}
                 - h b^{-2}\Box_n\varphi-(n-1)b^{-2}{\cal K}h\varphi \nnb\\
      &&\hspace{37ex}
                 +\frac12b^{-2}\sigma_{\pm}\kappa^2
                      \left(-\varphi\Box_n\varphi-n{\cal K}\varphi^2\right)
           \bigg]\,;
         \qquad
     \hat{k}_{\mu\nu}:=\frac12b^{2}\ppp_r\!\left(b^{-2}h_{\mu\nu}\right).
\label{actionSubs}
\end{eqnarray}
Note that $\varphi$ here is defined as the displacement of the 
corresponding
brane measured from the side of smaller values of $r$, that is,
\begin{eqnarray}
   \varphi |_{\Sigma_\pm}=\pm L(z_{\pm};\alpha) \tilde{\varphi}\,.
\end{eqnarray}

Here, it is worth noting that the above action $I^{(2)}$ is different
from the one obtained in \cite{GS}
(see Eq.~(A.19)).
This is because the above action $I^{(2)}$ is obtained by substituting a
particular form of the metric to the action of the system,
while the action of \cite{GS} is constructed without any restriction on the
metric. The particular form of the metric in Eq.~(\ref{result}) solves the
equation of motion in $z$-direction, and thus already extremize
the action in the $z$-direction.
Therefore, $I^{(2)}$ is an action to be maximized in the
$x^\mu$-directions only, which is to give the equation of motion
for $h_{\mu\nu}$ as a function of $\varphi(x^\mu)$.
Meanwhile, the action obtained in \cite{GS} is an action to be maximized
in both the $z-$direction and $x^\mu$-directions
independently to give the
equations of motion for both $h_{\mu\nu}$ and $\varphi$\cite{RMW}.

Let us define ${\cal L}_h$ and ${\cal L}_\varphi$ as
\begin{eqnarray}
  &&\quad
     {\cal L}_h:=-\hat{k}^{\mu\nu}h_{\mu\nu}+\hat{k}_{\sigma}^{\ \sigma}h
 +h_{\mu\nu}\varphi^{|\mu\nu}-h b^{-2}\Box\varphi -(n-1)b^{-2}{\cal 
K}h\varphi
  \nnb\\
  &&\quad
     {\cal L}_\varphi
       :=\frac12\Big(h_{\mu\nu}\varphi^{|\mu\nu}
                     -h b^{-2}\Box\varphi -(n-1)b^{-2}{\cal K}h\varphi\Big)
          +\frac12b^{-2}\sigma_{\pm}\kappa^2
                \bigg(-\varphi\Box\varphi-n{\cal K}\varphi^2\bigg)\,.
\end{eqnarray}
Then $I^{(2)}_{\Sigma_\pm}$ is express as
\begin{eqnarray}
   &&
   I^{(2)}_{\Sigma_\pm}
    =\frac{1}{\kappa^2}\int_{\Sigma_\pm}\hspace{-1ex} d^n x \sqrt{-q}
          \bigg[ \frac12{\cal L}_h +{\cal L}_\varphi \bigg]\,.\nnb
\end{eqnarray}
We find that the variation of ${\cal L}_h$ with respect to $h_{\mu\nu}$
gives the boundary condition of $h_{\mu\nu}$ on the brane, while that
of ${\cal L}_{\varphi}$ with respect to $\varphi$ gives the equation of
motion for $\varphi$.

With the substitution of Eq.~(\ref{result}) to $I^{(2)}_{\Sigma_\pm}$,
${\cal L}_h$ vanishes and only
${\cal L}_\varphi$ remains.
Using the commutation rule (\ref{commutation}),
${\cal L}_\varphi$ is simplified to give
\begin{eqnarray}
   I^{(2)}
     &=& I^{(2)}_{\Sigma+}+I^{(2)}_{\Sigma-}
      = \frac{(n-1)\ell^{n-2}}{\kappa^2}\int d^n\!x \sqrt{-\gamma}\,
           \Big(b_{-}\ppp_z u^{(2)}(z_-)
                      -b_{+}\ppp_z u^{(2)}(z_+)\Big)
                   \~\varphi\Big[(\Box+n{\cal K})\Big]\~\varphi\,,\nnb
\end{eqnarray}
which is Eq.~(\ref{pre-action}).

The effective action for the radion has been derived previously by
Chiba\cite{CH} and by Chacko and Fox\cite{CF} in the longitudinal
gauge. However, in our opinion, there seems to exist some
subtleties in their derivations that may be worth clarified,
apart from the difference in the choice of gauge.
We therefore compare their derivations with ours.

First, we need to transform the metric perturbation in
Eq.~(\ref{result}), which is given in the RS gauge,
to that in the longitudinal gauge in which the locations of the
branes are unperturbed, $h_{r\mu}=0$ and
 $h_{\mu\nu}\propto\gamma_{\mu\nu}$.
With the coordinate transformation of the form,
\begin{eqnarray}
   r \to r -L(z;\alpha)\~\varphi\,,\qquad
   x^\mu \to x^\mu -\{ \alpha\ell J+\psi^{(2)}_n\}\~\varphi^{|\mu}\,,
\end{eqnarray}
we obtain the metric perturbation in the desired gauge:
\begin{eqnarray}
   h_{rr}=-2(n-2)(b/\ell)^{2-n}\~\varphi\,,\quad
   h_{r\mu}=0\,,\quad
   h_{\mu\nu}=2b^2(b/\ell)^{2-n}\~\varphi\gamma_{\mu\nu}\,.
\label{LongGauge}
\end{eqnarray}

Let us first consider Chiba's effective action of radion given in
Eq.~(14) of \cite{CH}.
The metric ansatz Chiba uses to derive the action is
\begin{eqnarray}
  ds^2 = h_{,r}^2dr^2+e^{-2kh}\bar{g}_{\mu\nu}dx^\mu dx^\nu\,;
  \quad h=r+f(x^\mu)e^{2kr}\,,
\label{CGRmetric}
\end{eqnarray}
in which $\bar{g}_{\mu\nu}$ is a general 4-dimensional metric that
depends solely on $x^\mu$.
Here, the coordinate $z$ in his paper is replaced with the coordinate
$r$ to avoid confusion.
Linearizing the metric~(\ref{CGRmetric}) with respect to $f(x^\mu)$,
it reduces to the metric perturbation of the
form~(\ref{LongGauge}) with the
following correspondence between his notation and ours:
\begin{eqnarray}
   x^\mu \leftrightarrow \ell x^\mu\,,\quad
   k \leftrightarrow \ell^{-1}\,,\quad
   f(x^\mu) \leftrightarrow -\ell\~\varphi(x^\mu)\,,\quad
   \bar{g}_{\mu\nu} \leftrightarrow \gamma_{\mu\nu}\,,
\label{CHIBAcorresp1}
\end{eqnarray}
where we have fixed $n=4$. (It may be noted that Eq.~(\ref{CGRmetric})
can be extended to the general $n$ dimensions simply
by replacing $h=r+f(x^\mu)e^{2kr}$
with $r+f(x^\mu)e^{(n-2)kr}$ in the case of ${\cal K}=0$.)
With the above correspondence, we find his result,
when linearized, does not agree with ours.

Let us examine where the difference comes from by following his derivation.
Substituting the decomposition of the Ricci scalar of the
 metric~(\ref{CGRmetric}),
\begin{eqnarray}
   R[g]
     &=&-20k^2+e^{2kh}R[\bar{g}]
          +6k^2e^{4kr+2kh}h_{,r}^{-1}(-1+2kfe^{2kr})\V^\rho f \V_\rho f
          +(\mbox{total divergence term in $x^\mu$})\,,
\label{Rdecomp}
\end{eqnarray}
into his starting action of the system, Eq.~(1) of \cite{CH},
and integrating it with respect to $r$, we obtain
\begin{eqnarray}
 S_C
 &=&2\int d^4\!x\sqrt{-{g}_{(+)}}
       \bigg\{5kM^3(\alpha^{-2}e^{-4(\alpha-1)kf}-1)
             +\frac{M^3}{2k}
                \left(1-\frac{1}{\alpha}e^{-2(\alpha-1) kf}\right)
               e^{2kf} R[\bar{g}]\nnb\\
  & &\hspace{30ex}
             -3kM^3(\alpha-1)(\V_{(+)}f)^2
             -3kM^3(\alpha^{-2}e^{-4(\alpha-1)kf}-1)
       \bigg\}\nnb\\
  & &\hspace{14ex}
       -\sigma_{(+)}\int d^4\!x\sqrt{-{g}_{(+)}}
       -\sigma_{(-)}\int d^4\!x\sqrt{-{g}_{(-)}}\,,
\label{Cact}
\end{eqnarray}
where the following correspondence is understood:
\begin{eqnarray}
   g_{(\pm)\mu\nu} \leftrightarrow q_{\pm\mu\nu}\,,\quad
   M^3 \leftrightarrow \kappa^{-2}\,,\quad
   \alpha \leftrightarrow (\ell/b_{-})^{2}\,,\quad
   \sigma_{(\pm)} \leftrightarrow \sigma_{\pm}\,,\quad
   \Big(\Lambda \leftrightarrow \Lambda_{n+1}/(2\kappa^2)\Big)\,.
\label{CHIBAcorresp2}
\end{eqnarray}

Comparing the action (\ref{Cact}) with Eq.~(5) in \cite{CH},
it seems that the first term $5kM^3(\alpha^{-2}e^{-4(\alpha-1)kf}-1)$
is missing there, which comes from the integration of
the Ricci scalar of the background RS bulk spacetime, $-20k^2$ in
Eq.~(\ref{Rdecomp}).
Furthermore, if the correspondence (\ref{CHIBAcorresp2}) we
deduced is correct, the values of the brane tensions
$\sigma_{(\pm)}=\pm3M^3k$ as stated in Eq.~(3) of \cite{CH}
are wrong by the factor 2; i.e., the correct values are
$\sigma_{(\pm)}=\pm6kM^3$.
Besides, in his starting action where the integral with respect to
the extra dimension is given by twice the integral over $r$ in the
range $(0,r_c)$, that is, from the positive tension brane to the
negative tension brane, the geometrical boundary terms of
the bulk spacetime, $I_{K,1}$ and $I_{K,2}$ as in our
action of the system in Eq.~(\ref{basicAction}), should be added.

With these corrections we find
\begin{eqnarray}
&&
 S_{C}(\sigma_{(\pm)}\to\pm6kM^3)+I_{K,1}-I_{K,2}\nnb\\
&&\hspace{10ex}
  =2\int d^4\!x\sqrt{-{g}_{(+)}}
       \bigg\{\frac{M^3}{2k}
                \left(1-\frac{1}{\alpha}e^{-2(\alpha-1) kf}\right)
               e^{2kf} R[\bar{g}]
             -3kM^3(\alpha-1)(\V_{(+)}f)^2
       \bigg\}\,,\nnb
\end{eqnarray}
where we have used the fact that $K_{\mu\nu}=-k e^{-kh}\bar{g}_{\mu\nu}$.
Since our effective action of the radion is obtained by fixing
$\bar{g}_{\mu\nu}=\eta_{\mu\nu}$, the action to be compared is
\begin{eqnarray}
   S_{C}(\sigma_{(\pm)}\to\pm6kM^3,\bar{g}_{\mu\nu}\to\eta_{\mu\nu})
        +I_{K,1}-I_{K,2}
    =2\int d^4\!x\sqrt{-{g}_{(+)}}
       \bigg\{ -3kM^3(\alpha-1)(\V_{(+)}f)^2
       \bigg\}\,.
\label{FixedChibaAction}
\end{eqnarray}
Using Eqs.~(\ref{CHIBAcorresp1}) and (\ref{CHIBAcorresp2})
 and the correspondence
$N_{2{\cal K}}\leftrightarrow(\alpha-1)/2$ which follows from them,
we find that the above action agrees with our result.
%In passing,it should be noted that his final result in Eq.~(14) of
%\cite{CH} may be obtained by using the relation
%$e^{2kf} R[\bar{g}] =R_{(+)}-6k\Box_{(+)} f +6k^2(\V_{(+)}f)^2$
%which is manifest only when $\bar{g}_{\mu\nu}$ is not fixed.

Let us next consider the derivation by Chacko and Fox and their
result.
The correspondence between their notation and ours are
\begin{eqnarray}
&& G_{MN} \leftrightarrow g_{ab}\,,\quad
   \bar{G}_{\mu\nu} \leftrightarrow q_{\mu\nu}\,,\quad
  M^3 \leftrightarrow (4\kappa^2)^{-1},\quad
  \Lambda_B \leftrightarrow \Lambda_5/\kappa^2\,,\quad
  \bar{\Lambda}_0 \leftrightarrow \sigma_+\,,\quad
  \bar{\Lambda}_A \leftrightarrow \sigma_-\,,\quad
\nnb\\
&& \hspace{10ex}
  \alpha \leftrightarrow 2\ell^{-1}\,,\quad
  H \leftrightarrow 1\,,\quad f \leftrightarrow b^2\,,\quad
  \psi \leftrightarrow 2\ell^2\~\varphi\,,\quad
  g_{\mu\nu} \leftrightarrow \gamma_{\mu\nu}\,,
\end{eqnarray}
from which we deduce
\begin{eqnarray}
   \int^a_0 dr' f^{-1} \leftrightarrow \ell^{-1}N_{2{\cal K}}\,.
\end{eqnarray}
Using the above correspondence, we find that their result
given in Eq.~(34) of \cite{CF} exactly agrees with ours.

There is, however, a slight subtlety which may be worth
pointing out. It is the following.
Chacko and Fox state that they work in the compact extra dimension
with $Z_2$ symmetry. Therefore there is no need to
introduce the geometrical boundary terms in the action.
In fact, the form of the linearized action given by Eq.~(33) of
\cite{CF} is clearly free from such terms.
%In this setting, the curvature singularities at the branes contribute to
%the integral over $r$ of the linearized action.
However, if we perform the integral over $r$ of the linearized
action, there appear contributions from the branes due to the
curvature singularities.
The interesting fact is that the contributions cancel out with those
from the tension terms in the longitudinal gauge. 
Perhaps it is related with the fact that the coordinated
locations of the branes are unperturbed in the longitudinal gauge.
This implies that one may
arrive at the correct answer by simply ignoring the singular
contributions from the branes and simultaneously the tension terms, and
then integrating the linearized action only over the bulk 
between branes.
%in the end even if one
%would have carelessly ignored these singular contributions and
%integrated the linearized action only over the bulk
%between the branes.

In the present paper, we took a different approach.
That is, we divided
the covering space of the bulk into the two patches
bounded by the branes. This division requires
the additional geometrical boundary terms for each patch
to the action of the system.
If we apply our method to their starting action,
Eq.~(1) of \cite{CF}, it is rewritten as
\begin{eqnarray}
   S_{CF}
%&=&\int dx^4dr [\sqrt{-G}(2M^3R-\Lambda_B)
%                     -\delta(r)\sqrt{-\bar{G}}\bar{\Lambda}_0
%                     -\delta(r-a)\sqrt{-\bar{G}}\bar{\Lambda}_A]\nnb\\
    &=&2\int_{r=0}^{r=a}\hspace{-1ex} d^4x\sqrt{-G}(2M^3R-\Lambda_B)
       -\int_{r=0}\hspace{-2ex} d^4x\sqrt{-\bar{G}}\bar{\Lambda}_0
       -\int_{r=a}\hspace{-2ex} d^4x\sqrt{-\bar{G}}\bar{\Lambda}_A\nnb\\
    & &\hspace{8ex}
       +2\int_{r=0}\hspace{-2ex} d^4x\sqrt{-\bar{G}}\, (4M^3 K)
       -2\int_{r=a}\hspace{-2ex} d^4x\sqrt{-\bar{G}}\, (4M^3 K)\,,
\label{(1)rw}
\end{eqnarray}
where $K_{\mu\nu}$ is the extrinsic curvature of the $r$-constant
hypersurface defined with the normal vector pointing toward the
increasing direction of $r$.
Apart from the difference in notation as expressed in the above,
this form of the action is the one we used as the starting point
of our calculation. See Eq.~(\ref{basicAction}).
Substituting the perturbation given in Eq.~(\ref{LongGauge}) into the
above action, 
we find that the terms arising from the first integral
in Eq.~(\ref{(1)rw}) that can be cast into the total divergence
form cancel out with the perturbation of the last four terms of
Eq.~(\ref{(1)rw}), i.e., the tension terms and the geometrical boundary
terms. This cancellation gives the same form of action as the one in
their Eq.~(33) by restricting the integral only over the bulk,
and consequently leads to their final result, Eq.~(34).

\section{Effective Brans-Dicke Parameter}\label{BDparameter}

In this appendix, we derive the effective Brans-Dicke parameter on the
$n$-dimensional positive tension brane with curvature ${\cal K}=0$, $1$.

The action for the Brans-Dicke theory in the $n$-dimensional space-time is
\begin{eqnarray}
  I=\frac{1}{16\pi G_n}\int \sqrt{-q}\,d^n\!x
      \left[ \Psi ({}^{(n)}R-2\Lambda) -\Psi^{-1}\omega_{\rm BD}\,
                      q^{\mu\nu}\ppp_{\mu}\Psi\ppp_{\nu}\Psi
            \right]
    +\int \sqrt{-q}\,d^n\!x{\cal L}_{m}
\end{eqnarray}
where $q_{\mu\nu}$ is the metric of the n-dimensional space-time,
$\omega_{\rm BD}$ is the Brans-Dicke parameter,
and ${\cal L}_m$ is the Lagradian density of matter fields.
Let $\Psi(x^{\mu})=f_0 \exp[ W(x^\mu)]$. Then, the equations of motion
linearized in $W(x^\mu)$ are
\begin{eqnarray}
  && {}^{(n)}G_{\mu\nu}+\Lambda q_{\mu\nu}
             =\frac{8\pi G_n}{f_0}\, \tau_{\mu\nu}
                 +\left[D_{\mu}D_{\nu}-q_{\mu\nu}\Box_q\right] W(x^\mu)\,,
\label{EinsteinBD}\\
  && \hspace{2ex}
       f_0 \Box_q W(x^\mu)=\frac{1}{n-1+(n-2)\omega_{\rm BD}}
                              \left(8\pi G_n\tau-2\Lambda f_0\right)\,,
\label{EomBD}
\end{eqnarray}
where $\tau_{\mu\nu}$ is the energy-momentum tensor that arises from
${\cal L}_m$.
Using Eq.~(\ref{EomBD}), Eq.~(\ref{EinsteinBD}) can be put into the
following form:
\begin{eqnarray}
 &&
   {}^{(n)}G_{\mu\nu}
      +\frac{n-2}{n}\frac{n-1+n\,\omega_{\rm BD}}{n-1+(n-2)\omega_{\rm BD}}
             \Lambda\, q_{\mu\nu}\nnb\\
 &&\hspace{10ex}
     =\frac{8\pi G_n}{f_0}\,
       \frac{(n-2)\omega_{\rm BD}}{n-1+(n-2)\omega_{\rm BD}}\tau_{\mu\nu}
        +\frac{8\pi G_n}{f_0}\,\frac{n-1}{n-1+(n-2)\omega_{\rm BD}}
                  \bar{\tau}_{\mu\nu}
        + \left[D_{\mu}D_{\nu}-\frac1{n} q_{\mu\nu}\Box_q\right] W(x^\mu)
  \label{BD-Einstein}
\end{eqnarray}
where $\bar{\tau}_{\mu\nu}:=\tau_{\mu\nu}-(1/n)q_{\mu\nu}\tau$.

To obtain the effective Einstein equations on the positive tension brane
for the energy density small compared with the
tension of the brane, we substitute Eq.~(3.8) of \cite{GS} into
Eq.~(\ref{branEin}) of this paper.
The result is
\begin{eqnarray}
  {}^{(n)}G_{\mu\nu}+\Lambda_n q_{\mu\nu}=\kappa_n^2\tau_{\mu\nu}
  +\frac{(\ell/b_+)^{n-2}\kappa^2}{2\ell N_{(n-2){\cal 
K}}}\bar\tau_{\mu\nu}
      +\frac{(\ell/b_+)^{n-2}}{\ell N_{(n-2){\cal K}}}
          \left[D_{\mu}D_{\nu}-\frac1{n} q_{\mu\nu}\Box_q\right] \phi\,.
  \label{EffectiveEinstein}
\end{eqnarray}

Comparing Eq.~(\ref{BD-Einstein}) with Eq.~(\ref{EffectiveEinstein}), we
find that the effective gravity on the brane takes the Brans-Dicke form
with the following identifications:
\begin{eqnarray}
 && \kappa_n^2
       =\frac{8\pi G_n}{f_0}\,
               \frac{(n-2)\omega_{\rm BD}}{n-1+(n-2)\omega_{\rm BD}}\,,
       \qquad
    \frac{(\ell/b_+)^{n-2}\kappa^2}{2\ell N_{(n-2){\cal K}}}
      =\frac{8\pi G_n}{f_0}\,\frac{n-1}{n-1+(n-2)\omega_{\rm BD}}
\label{omega}\\
 && \hspace{25ex}
  \frac{(\ell/b_+)^{n-2}}{\ell N_{(n-2){\cal K}}}\phi=W(x^{\mu})\,.
\end{eqnarray}
Eliminating $f_0$ from Eq.~(\ref{omega}), we obtain
\begin{eqnarray}
  \omega_{\rm BD}=(n-1)\frac{\cosh\sqrt{\cal K}z_+}
                            {(\sinh\sqrt{\cal K}z_+/\sqrt{\cal K})^{n-2}}
                  N_{(n-2){\cal K}}\,,
\label{BDpara}
\end{eqnarray}
while eliminating $\omega_{\rm BD}$ from
Eq.~(\ref{omega}), we obtain
\begin{eqnarray}
  8\pi G_n=\frac{\kappa^2\ell^{-1}}{2}
      \left[ (n-2)\cosh\sqrt{\cal K}z_+
            +\frac{1}{N_{(n-2){\cal K}}}
               \Big(\sinh\sqrt{\cal K} z_+/\sqrt{\cal K}\Big)^{n-2}
     \right]
\label{Gn}
\end{eqnarray}
For ${\cal K}=1$ and $n=4$, we have
$N_{(n-2){\cal K}}=\cosh z_{-}-\cosh z_{+}$ and
\begin{eqnarray}
&&
 \omega_{\rm BD}
     = 3\frac{\cosh z_+}{\sinh^2 z_+} N_{2{\cal K}}\,,
%    = 3\frac{\sqrt{1+(\ell/H_+^{-1})^2}}{(\ell/H_+^{-1})^2}N_{2{\cal K}}
\qquad
%\\&&
  8\pi G_4=
%   \left\{
%     \begin{array}{ll}
%     \displaystyle
       \frac{\kappa^2\ell^{-1}}{2}
          \left[ 2\cosh z_+ +\frac{\sinh^2 z_+}{N_{2{\cal K}}}
         \right]\,.
%       =\frac{\kappa^2\ell^{-1}}{2}
%          \left[ 2\sqrt{1+(\ell/H_+^{-1})^2}
%                +\frac{(\ell/H_+^{-1})^2}{N_{2{\cal K}}}
%         \right] \quad
%       & \Big({\cal K}=1\Big)\\
%     \displaystyle
%       \kappa^2\ell^{-1}\frac{(z_-/z_+)^2}{(z_-/z_+)^2-1}
%       & \Big({\cal K}=0\Big)
%     \end{array}
%   \right.
\label{BDparaGnK=1}
\end{eqnarray}
For ${\cal K}=0$ and $n=4$, we have $N_{(n-2){\cal K}}=(z_-^2-z_+^2)/2$ and
Eqs.~(\ref{BDpara}) and (\ref{Gn}) recover the result of Garriga and
Tanaka\cite{GT}.

\section{$\bm E_{\mu\nu}$ \`{a} la
cosmological perturbation theory}\label{DecayEmn}

In this appendix, we express the effect of the radion on the brane
in the language of cosmological perturbation theory.
We follow the notation of \cite{KS}.
To begin with, we list a few important facts to be kept in mind:
\begin{list}{}{}
\item[(a)] The background is de Sitter space-time.
\item[(b)] $\tau^E_{\mu\nu}=-E_{\mu\nu}/\kappa_n^2$ is gauge-invariant
 as it is.
\item[(c)] $\tau^E_{\mu\nu}$ is traceless.
\end{list}

The points (a) and (b) are somewhat related.
As long as the background bulk spacetime is anti-de Sitter,
the point (b) is always true.
Independent of this fact, from the brane point
of view, the matter perturbation variables are automatically
gauge-invariant on the pure de Sitter space.
For example, if we consider the density perturbation,
the gauge transformation law for $\delta\rho$ is given by
\begin{eqnarray}
\delta\rho\to \overline{\delta\rho}
=\delta\rho+(n-1)aH(\rho+p)T\,,
\end{eqnarray}
where $T$ is the gauge function describing the shift in the
time coordinate; $\eta\to\bar\eta=\eta+T$.
Thus the density perturbation
on exactly de Sitter space where $\rho+p=0$ is always gauge-invariant,
and the same is true for all the matter perturbation variables.
It should be noted, however, this special property of
the de Sitter background can lead to an erroneous result
if one tries to consider the perturbation in
a gauge specified by conditions involving the matter variables.
We shall come back to this issue later.
For the moment, what we can claim is that
the gauge-invariance is guaranteed provided
one considers the perturbations in a class of gauges
whose conditions involve only the geometrical variables
such as the metric. Let us call this class of gauges
the geometrical gauges.

First we introduce the matter perturbation variables:
\begin{eqnarray}
  && \delta T^0{}_0=-\rho\,\delta(\eta)\,Y(x^i)\,,
\quad
  \delta{T}^{0}{}_{j}=\rho\,q(\eta)Y_j(x^i)\,,
\nnb\\
&&
\delta{T}^i{}_j=P\,\pi_L(\eta) \delta^i_j\,Y(x^i)
      +P\pi_T(\eta) Y^i{}_{j}(x^i)\,,
\label{DefTotalEM}
\end{eqnarray}
where $Y(x^i)$, $Y_i(x^i)$ and $Y_{ij}(x^i)$ are the spatial
harmonics satisfying
\begin{eqnarray}
    ({\cal D}^k{\cal D}_k+k^2)Y(x^i)=0\,,\quad
    Y_i(x^i):=-k^{-1}{\cal D}_iY(x^i)\,,\quad
    Y_{ij}(x^i):=k^{-2}
       \left[{\cal D}_{i}{\cal D}_{j}
               -\frac{1}{3}\sigma_{ij}{\cal D}^k{\cal D}_k\right]Y(x^i)\,,
\end{eqnarray}
and $\delta{T}^{\mu}{}_{\nu}$ is just the effective radion
energy momentum tensor,
\begin{eqnarray}
   \delta{T}_{\mu\nu}=\tau^E_{\mu\nu}=-{1\over\kappa_n^2}E_{\mu\nu}\,.
\end{eqnarray}
Note that the usual velocity perturbation, $v=\rho q/(\rho+P)$,
cannot be defined because of $\rho+P=0$ on the background.
Also note that the traceless nature of $\tau^E_{\mu\nu}$
implies $\pi_L=-\delta/(n-1)$.
As for the metric perturbation, we introduce the variables as
\begin{eqnarray}
h_{\mu\nu}dx^\mu dx^\nu
= - 2 a^2 A Y d\eta^2- 2 a^2 B Y_jd\eta dx^j
           +a^2\left[ 2H_L\delta_{ij}+2H_TY_{ij}\right]dx^i dx^j\,,
\end{eqnarray}
where we have chosen the spatially flat chart of the de Sitter space.

As noted above, the choice of gauge is irrelevant as long as we
choose a geometrical gauge. The perturbation equations are then
derived by simply writing down
the energy momentum conservation law $\delta T^{\mu\nu}{}_{;\nu}=0$
on the unperturbed background.
Therefore, for definiteness, let us choose
the Newton (or longitudinal) gauge in which
the shear of the constant time hypersurfaces vanishes,
i.e., $k^{-1}H_T'-B=0$,
and denote the matter variables in this gauge as
\begin{eqnarray}
\delta=-(n-1)\pi_L=\Delta_s\,,\quad
q=Q_s\,,\quad
 \pi_T=\Pi\,.
\end{eqnarray}
Adopting the scale factor $a$ as the time variable, we find
\begin{eqnarray}
&&a{d\over da}\Delta_s+n\Delta_s=-{k\over aH}Q_s\,,
\label{Deltaseq}\\
&&a{d\over da}Q_s+nQ_s={k\over aH}\left(
{1\over n-1}\Delta_s+{n-2\over n-1}\Pi\right)\,.
\label{Qseq}
\end{eqnarray}
These equations can be combined to give a second order
equation for $\Delta_s$,
\begin{eqnarray}
   a^2\frac{d^2}{da^2}\Delta_{\rm s}
     +2(n+1)a\frac{d}{da}\Delta_{\rm s}
       +\left(\frac{1}{n-1}\frac{k^2}{a^2H^2}+n(n+1)\right)
            \Delta_{\rm s}
    = -\frac{n-2}{n-1}\frac{k^2}{a^2H^2} \Pi.
\label{EvolDelta2}
\end{eqnarray}

We see from the above equation that if there were no anisotropic
stress $\Pi$, then $\Delta_{\rm s}\propto a^{-n}$ or $a^{-(n+1)}$
after horizon-crossing.
However, in the present case, we have
$\Pi\propto k^2a^{-2}\psi\propto ka^{-1}$.
This slow decay rate of the anisotropic stress acts as a
source to the energy density $\Delta_{\rm s}$. With $\Pi\propto ka^{-1}$,
the right hand side of Eq.~(\ref{EvolDelta2}) behaves as $k^3a^{-5}$,
which implies $\Delta_{\rm s}\propto k^3a^{-3}$ (for $n>3$).
The behavior of $Q_s$ is found as $Q_s\propto k^2a^{-2}$.
To be a bit more precise, the leading order behaviors of
$\Delta_s$, $Q_s$ and $\Pi$ are given by
\begin{eqnarray}
\Delta_s=-{C\over(n-1)(n-3)}\left({k\over aH}\right)^3\,,
\quad
Q_s={C\over n-1}\left({k\over aH}\right)^2\,,
\quad
\Pi=C\left({k\over aH}\right)\,,
\label{leading}
\end{eqnarray}
where $C$ is a constant.
These results are of course fully consistent with those obtained
in the main text, Eq.~(\ref{Emnform}).
Thus one may say that this unusual behavior of $\Delta_s$
is caused by the large (though decaying) anisotropic stress.

It is instructive to re-express the above result in the
so-called comoving gauge in which the $\eta=$constant
hypersurfaces are chosen in such a way that
\begin{eqnarray}
\delta{T}^0{}_j=\rho\, q\,Y_j=0\,.
\label{comoving}
\end{eqnarray}
Note that this condition involves the matter variable $q$.
A peculiarity of the comoving gauge is that
the density perturbation in this gauge
is not equal to $\Delta_s$, although the matter perturbation variables
should be gauge-invariant on the pure de Sitter space-time as noted above.
The density perturbation on the comoving hypersurface, which
we denote by $\Delta$, is related to $\Delta_s$ as \cite{KS}
\begin{eqnarray}
   \Delta=\Delta_{s}+(n-1){aH\over k}Q_s\,.
\label{defDelta}
\end{eqnarray}
One immediately sees from Eq.~(\ref{leading}) that
the $Q_s$ term dominates and the leading order behavior of
$\Delta$ is given by
\begin{eqnarray}
\Delta=C{k\over aH}\,,
\label{Dlead}
\end{eqnarray}
which decays much slower than $\Delta_s$.
The cause of this seemingly inconsistent result is
the comoving gauge condition (\ref{comoving})
which forces $q$ to be zero. 
As $q$ is a gauge-invariant quantity, this condition would never be
fulfilled and thus the comoving slice does not exist.
In fact, it is straightforward to show that all the metric
perturbation variables are ill-defined in the comoving gauge. 
Therefore, although the variable $\Delta$ is well-defined
as in Eq.~(\ref{defDelta}), this $\Delta$
does not represent a density perturbation in any gauge. 

Let us analyze the behaviors of the metric perturbation
variables in the Newton gauge.
Although not essential, for simplicity, we take $H_T=0$ which
implies $B=0$.
We denote the metric variables in the Newton gauge by
\begin{eqnarray}
A=\Psi\,,\quad H_L=\Phi\,,
\end{eqnarray}
Then the $(0,\mu)$-components and the traceless part of
the $(i,j)$-components of the Einstein equations give\cite{KS}
\begin{eqnarray}
&&(n-2){k^2\over a^2}\Phi=\kappa_n^2\,\rho\,\Delta
=3H^2\Delta\,,
\nnb\\
&&(n-3)\Phi+\Psi=-\kappa_n^2{a^2\over k^2}P\Pi
=3\left({aH\over k}\right)^2\Pi\,.
\end{eqnarray}
Inserting the leading behaviors of $\Delta$ and $\Pi$ given above
to these equations, we find
\begin{eqnarray}
\Psi=\Phi={3\over n-2}\left({aH\over k}\right)\,.
\end{eqnarray}
One may find it a bit surprising that these variables grow as $a$,
seemingly indicating an instability. The resolution is that the Newton
gauge is not really a good gauge. One should examine if
there is a different choice of gauge in which the metric perturbations
behave regularly, and there exists indeed such a gauge.
By the shift of the time slice $\eta\to\bar\eta=\eta+T$,
$A$ and $H_L$ transform as
\begin{eqnarray}
A\to\bar A=A-H{d\over da}(aT)\,,
\quad
H_L\to \bar H_L=H_L-HaT\,.
\end{eqnarray}
Then it is easy to see that the leading terms of $\Phi$ and $\Psi$
are simultaneously eliminated by $T=\Phi/(aH)=$constant.
Thus the above apparent instability is just
a reflection of the bad choice of gauge.

%======================================%
%<<<<<<<<<<<< REFERENCES >>>>>>>>>>>>>>%
%======================================%
\newcommand{\np}{Nucl. Phys. }

\end{document}